\documentclass[twocolumn,aps,prl,superscriptaddress,showpacs,floatfix]{revtex4}
\usepackage{amsmath,bm}
\usepackage{graphicx}
\usepackage{epsfig}
\begin{document}

\title{Finding the remnants of lost jets at RHIC}
\author{Subrata Pal}
\affiliation{National Superconducting Cyclotron Laboratory
and Department of Physics and Astronomy, Michigan State
University, East Lansing, Michigan 48824}
\author{Scott Pratt}
\affiliation{Department of Physics and Astronomy, Michigan State
University, East Lansing, Michigan 48824}

\begin{abstract}
A fast parton propagating through partonic matter loses energy by
medium-induced gluon radiation. We propose a method to detect the energy loss
of jets and the final-state interaction of the radiated gluons and the jets
with the partonic medium. By embedding jets and radiated energy into a parton
cascade, we find that the fate of lost jets can be studied by analyzing the
momentum distribution of soft hadrons relative to the jet axis.

\end{abstract}
\pacs{12.38.Mh, 24.85.+p, 25.75.-q}
\maketitle

The attenuation of hard jets produced in pQCD processes provides an excellent
opportunity to study the properties of hot and dense partonic matter expected
to be formed in ultrarelativistic heavy ion collisions
\cite{wang94,baier,wied,gyul}.  Due to the energy lost by the jets as a result
of multiple elastic and inelastic interactions in the partonic matter, the
final hadron $p_T$ distribution from the fragmenting jets will be suppressed
compared to that from nucleon-nucleon reference spectra convoluted with the
number of binary collisions. Measurement of high $p_T$ hadrons is thus expected
to provide a diagnostic tool for the total energy loss $\Delta E$ of jets in
the dense partonic medium \cite{wang02,hira}. Relative to the binary collision
scaled $p\bar p$ reference spectrum, experimental data
\cite{phenix,star,phobos,brahms} for central Au+Au collisions at RHIC 
indeed showed a suppression for $\pi^0$s and $h^\pm$, where the suppression is
smallest at $p_T \approx 2$ GeV/c and increases to an approximately constant
value for $p_T = 4-10$ GeV/c.  Apart from the suppression of high $p_T$ hadrons
at RHIC, the suppression of the away-side jets compared to the near-side jets
at high $p_T$
\cite{star2} is an indication of jet quenching for the away-side jet.

Even the disappearance of away-side jets might be considered as only
circumstantial evidence of jet-energy loss. For instance, one could imagine a
scenario where mono-jets are produced due to the $2\to 1$ gluon fusion processes,
e.g., in the gluon saturation model \cite{kharzeev}. If the away-side 
jet were truly produced then absorbed by the medium, the extra momentum and 
energy should survive. Our aim is to investigate the degree to which the lost 
energy and momentum can be extracted from low-$p_T$ observables. We propose a
straight-forward strategy of gating on the observation of a high-energy parton,
then analyzing the accompanying alteration of the low-energy spectra relative
to the azimuthal direction of the jet. The distribution of these extra
particles, as a function of $p_T$ and as function of their rapidity and
azimuthal angle relative to the triggering jet, might provide critical insight
for understanding the mechanism for jet-energy loss.

To investigate these issues, we consider a baseline dynamical model of the
collision, onto which hard jets have been embedded. The base
distribution consists of the HIJING model \cite{hijing,ampt2}, 
which is used to generate a set of initial partons, and ZPC, a parton cascade 
which simulates the subsequent elastic partonic interactions\cite{zpc}. 
HIJING provides the initial momentum and space-time coordinates of
partons from minijets and strings which originate from soft nucleon-nucleon
collisions. The parton-parton elastic scattering in ZPC is determined by the
cross section,
\begin{equation}\label{ppcrs}
\frac{d\sigma}{d\hat t} = C_a 4\pi\alpha^2_s \left( 1 + \frac{\mu^2}{\hat s}
\right) \frac{1}{(\hat t - \mu^2)^2} ~,
\end{equation}
where $\hat t$ and $\hat s$ are the usual Mandelstam variables, $C_a
=9/8,1/2,2/9$ for $gg, gq, qq$ scatterings, and $\alpha_s$ is the strong
coupling constant assumed to be 0.30.  The effective screening mass $\mu$ is
assumed to independent of temperature and density and is used as a parameter to
obtain the desired parton-parton elastic cross section $\sigma \approx C_a
4\pi\alpha^2_s/\mu^2$. We have used a typical value for $\mu = 0.50$ GeV. After
the partons freeze out, they are traced back to their parent strings and are
converted to hadrons using the Lund string fragmentation model \cite{lund}.

Since the jet production cross section is small, we include additional hard
jets in order to enhance statistics.  The multiplicity of the additional jets
in an event is obtained from a Poisson distribution with a mean $\langle
n\rangle = 10$.  The underlying dynamics of the base distribution ($dN_{\rm
parton}/dy\approx 250$ at $\sqrt s =200A$ GeV) is not significantly affected by
the smaller number of the embedded jets.  This technique of oversampling has
been widely used to study rare particle production, such as $K^-$ at SIS/GSI
\cite{kaon}, multistrange particles at SPS \cite{mstrange}, and $J/\psi$ at
RHIC \cite{jpsi}. In $p+p$ collision, the lowest order (LO) pQCD momentum
distribution of hard partons that propagate in the parton cascade is given by
\begin{eqnarray}\label{pjet}
E\frac{d\sigma^{pp}_j}{d^3p} &=& K\sum_{a,b} \int dx_a dx_b 
\int d^2{\bf k}_{T_a}
d^2{\bf k}_{T_b}  \: g({\bf k}_{T_a}) g({\bf k}_{T_b}) \nonumber \\ 
&&\times f_a(x_a,Q^2) f_b\left(x_b,Q^2\right) \: 
E\frac{d\sigma_{ab}}{d^3p} ~.
\end{eqnarray}
In Eq. (\ref{pjet}), $x_a$, $x_b$ are the momenta fractions and ${\bf
k}_{T_a}$, ${\bf k}_{T_b}$ are the intrinsic transverse momenta of the initial
scattered partons. A constant factor $K=2$ is used to account for the NLO
corrections. For the collinear parton distribution function (PDF)
$f_a\left(x_a,Q^2\right)$, we employ the LO Gl\"uck-Reya-Vogt (GRV94)
parametrization \cite{grv94}, and for the transverse PDF $g({\bf k}_{T_a})$, we
use Gaussian intrinsic $k_T$ distribution of width $\langle {\bf
k}_{T_a}^2\rangle = 1$ GeV$^2$/c$^2$. We neglect nuclear shadowing for the jets
as its effect is rather small at mid-rapidity at RHIC energies
\cite{jeon,klein}.  Since we explore events with at least one triggered hadron
of $p_T^{\rm trig} > 4$ GeV/c, a minimum momentum transfer of $p_T^{\rm min} =
4$ GeV/c for the additional jets is assumed. In the hard processes, a proper
formation time of $\tau_0 = 1/m_T$ is used for parton production with
transverse mass $m_T$. The initial transverse coordinate of a hard jet is
specified by the number of binary collision distribution for two Woods-Saxon
geometries.

The hard jets can suffer energy loss by elastic scattering with the partons
from HIJING.  However, the dominant energy loss is via medium-induced gluon
radiation along the trajectory of the jet. We consider here the dominant
first-order total energy loss expression in opacity, $\chi = L/\lambda_g$, from
the reaction operator approach \cite{gyul,vitev}:
\begin{eqnarray}\label{elos}
\Delta E = 2\pi\alpha_s^3 C_a C_A \int^\infty_{\tau_0} d\tau \rho({\bf r}(\tau),\tau)
(\tau - \tau_0) \log \frac{2E}{\mu^2 L} ~.
\end{eqnarray}
Here, ${\bf r}(\tau)$ and $E$ are the position and initial energy of the jet,
$\rho({\bf r}(\tau),\tau)$ is the parton density. The color factor $C_a$ is 
given in Eq. (\ref{ppcrs}) and may be expressed as $C_a =C_R C_T/d_A$, where 
$C_R$ and $C_T$ are the color Casimirs of the jet and the target parton, and 
$d_A$ is the dimension
in the adjoint representation with Casimir $C_A$. The effective path length
required by the radiated gluons for traversing the matter, $L$, is taken to be
4 fm.  Because of jet quenching the fragmentation function (FF) for a jet $c$
to a hadron $h$ in vacuum, $D_{h/c}(z_c,Q^2_c)$, is altered due to the
modification of the kinetic variables of the jet, and the effective
fragmentation function follows the relation \cite{gyul},
\begin{eqnarray}\label{ffn}
&\!\!\!\!& z_c D_{h/c}^\prime (z_c,Q^2_c) = z_c^\prime D_{h/c}(z_c^\prime,Q^2_{c^\prime}) 
+ N_g z_g D_{h/g}(z_g,Q^2_g) ~, \nonumber \\
&\!\!\!\!& z_c=\frac{p_h}{p_c}, ~~ z_c^\prime = \frac{p_h}{p_c-\Delta E(p_c)}, ~~
z_g = \frac{p_h}{\Delta E(p_c)/N_g} ~.
\end{eqnarray}
The second term on the right eventually gives the soft hadrons from the
fragmentation of $N_g$ radiated gluons. Note that the modified fragmentation
function satisfies the sum rule $\int dz_c \: z_c D_{h/c}^\prime (z_c,Q^2_c) =
1$. The initial hard jets after the partonic phase are fragmented using the LO
Binnewies-Kniehl-Kramer (BKK) parametrization \cite{bkk} of the fragmentation
functions.

In absence of any definitive relation for how radiated energy should be
translated into particles, we take recourse to a statistical
prescription. Assuming local thermodynamical equilibrium of the parton medium,
if $\Delta E({\bf r}, \Delta\tau)$ is the energy radiated in a time interval
$\Delta\tau$ (evaluated from Eq. (\ref{elos})), the number of thermal gluons
emitted in $\Delta\tau$ can be related to the entropy increase $\Delta S$ of
the plasma,
\begin{eqnarray}\label{ngluon}
N_g({\bf r}, \Delta\tau) \approx \frac{1}{4}\Delta S = 
\frac{1}{4} \frac{\Delta E({\bf r},\Delta\tau)} {T({\bf r},\tau)} ~,
\end{eqnarray}
where the local temperature $T \approx \epsilon({\bf r},\tau) /3\rho({\bf r},
\tau)$ is extracted from the cascade by viewing the local parton number and
energy densities $\rho({\bf r}, \tau)$ and $\epsilon({\bf r}, \tau)$. The
factor of $1/4$ comes from considering an ideal massless gas of classical
particles in equilibrium. The radiated gluons are formed at space-time points
along the parent jet's trajectory consistent with the energy loss described in
Eq. (\ref{elos}). They are initially directed along the axis of the parent jet
and subsequently scatter via elastic collision with the neighboring partons
after a formation time $\tau_0$ as used in HIJING. Effectively, this procedure
converts 100\% of the radiated energy into thermal energy, while maintaining
momentum conservation. A less-efficient prescription for thermalizing the
radiated energy would result in a reduced multiplicity. Based on local
parton-hadron duality \cite{eskola,molnar}, each of these radiated gluons after
freeze-out are converted to a pion with equal probability for the three charge
states.

\bigskip

\begin{figure}[ht]
\centerline{\epsfig{file=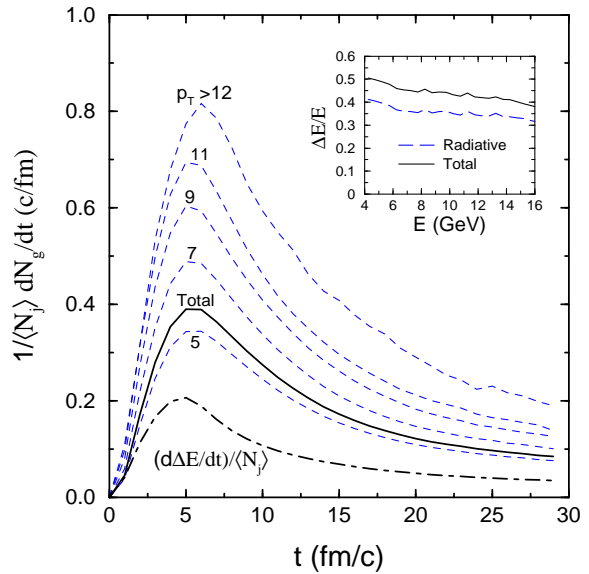,width=3.0in,height=3.0in,angle=0}}
\caption{ The production rate of gluons at all rapidities from radiative jet 
energy loss in central ($b=0$ fm) Au+Au collisions at RHIC c.m. energy of 
$\sqrt s = 200A$ GeV. The results are for jets with different $p_T$ (dashed lines) 
and for the total jets (solid line) in an event, normalized to the average 
number of jets $\langle N_j\rangle$ with the given $p_T$. The dash-dotted line 
gives the rate of radiative energy loss $(d\Delta E/dt)/\langle N_j\rangle$ in GeV/fm.  
The rise at small times is the result of performing the calculation in Cartesian 
time, rather than proper time. The inset shows the energy dependence of fractional 
energy loss $\Delta E/E$ of gluon jets due to radiation (dashed line) and after 
elastic scattering (solid).} 

\label{rate}
\end{figure}

The inset of Fig. \ref{rate} shows the fractional energy loss $\Delta E/E$ for
the fast gluons traversing the partonic medium for central Au+Au collisions at
a center of mass energy $\sqrt s = 200A$ GeV. In contrast to the asymptotic
BDMS \cite{baier} energy loss for a thick static plasma, $\Delta E_{\rm BDMS} =
\pi\alpha_s^3 C_R C_A/2 \cdot \tilde{v} \cdot L^2\rho$ (the factor $\tilde{v}
\propto \log(L/\lambda_g) \sim 1-3$), where $\Delta E/E$ increases rapidly with
decreasing energy, the GLV \cite{gyul,vitev} first-order radiative energy loss
$\Delta E$ (dashed line) exhibits almost a linear energy dependence for $E=4-7$
GeV while at higher energy $\Delta E/E$ reveals a $\log(E)/E$ decrease. Elastic
collision with other partons, which is modeled with the parton cascade, leads
to an additional energy loss by $25\%$ (solid line).

The dynamical evolution of the average energy loss per jet in an event
(for jets at all rapidities),
$(d\Delta E/dt)/\langle N_j \rangle$, rises for small times due to the
longitudinal expansion of the system as shown in Fig. \ref{rate}. If proper time
had been used rather than Cartesian time, the rate would have monotonically
fallen due to the falling density.  

Figure \ref{rate} also shows the rapidity integrated dynamical rate of gluon 
radiation per average
number of jets, $(dN_g/dt)/\langle N_j \rangle$, as evaluated from
Eq. (\ref{ngluon}). The results are for total jets in an event, and also for
jets with definite $p_T$. The emission pattern is found to follow that for the
energy loss. Though the number of radiated gluons per $\langle N_j\rangle$ is
seen to increase with $p_T$, the rapidly decreasing jet production cross
section with increasing jet energy $E$ actually leads to smallest number of
gluon contribution from jets with $p_T \geq 12$ GeV/c. Although higher-energy
jets are rare, they radiate more particles due to fact that they have more
available energy and due to the logarithmic dependence of the radiation in
Eq. (\ref{elos}). We find that partons with initial $p_T \sim 9$ GeV/c produce
the bulk of the radiated gluons. Using the statistical prescription described
above, about 5.5 soft gluons are emitted in an event per jet.

Since jets are typically formed in pairs, the observation of a jet can be used
as a trigger. One can then search for the remains of the balancing jet, as well
as any particle whose existence is derived from medium-induced radiation from the
triggered jet. For our calculations, the trigger is defined as a hadron at
midrapidity $|y|<1$ with its transverse momentum $p_T^{\rm trig}$ above 4
GeV/c.  The $x$ direction is then defined to be perpendicular to the beam axis
and pointing along the direction of the triggering hadron. Particles radiated from
the trigger jet would typically have positive values of $p_x$ while particles
associated with the jet's partner would have negative values of $p_x$. For a
perfect detector, momentum conservation implies
\begin{equation}
\label{eq:pxconstraint}
\int p_x \rho(p_x) \: dp_x = -P_T ~,
\end{equation}
where $P_T \equiv p_T^{\rm trig}$ is the momentum of the trigger particle and 
$\rho(p_x)=dN/dp_x$ refers to the momentum density of all other particles. 
Of course, this constraint is significantly relaxed by finite acceptance. 
By making a similar distribution $\rho_0$ from other events, without the 
requirement of a trigger, one can define
\begin{equation}
\Delta \rho(p_x)=\rho(p_x)-\rho_0(p_x) ~,
\end{equation}
which describes the variation of the momentum density due to the jet. Two
quantities related to $\Delta\rho$ are of particular interest:
\begin{eqnarray}\label{constr2}
\Delta N&\equiv&\int \Delta \rho(p_x) \: dp_x ~, \nonumber\\
\Delta P_x&\equiv&\int p_x\Delta \rho(p_x)  \: dp_x ~.
\end{eqnarray}
$\Delta N$ counts the net number of additional particles, while $\Delta P_x$
describes the momentum, which should be a fraction of $-P_T$, determined by the
acceptance. 

\bigskip

\begin{figure}[ht]
\centerline{\epsfig{file=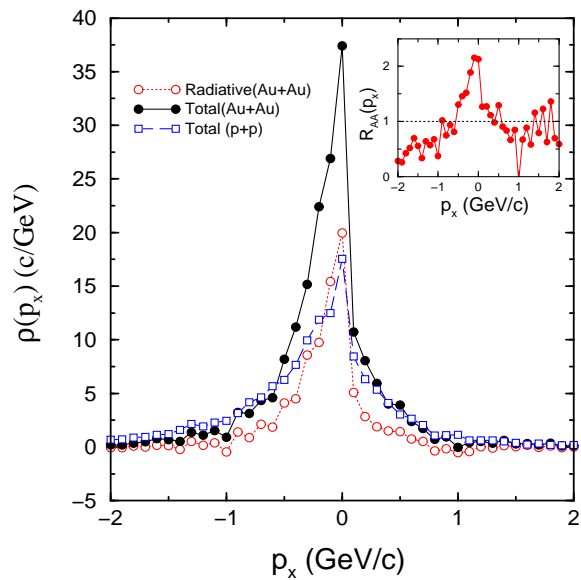,width=3.0in,height=3.0in,angle=0}}
\caption{The distribution $\rho(p_x)$ for particles produced at mid-rapidity
$|y|<1$ in events with a triggered hadron $4 < p_T^{\rm trig} < 6$ GeV/c
in the same rapidity range.
The results are for total number of particles (solid circles) and for
particles from radiative jet energy loss (open circles) in central ($b=0$ fm)
Au+Au collisions at RHIC energy of $\sqrt s = 200A$ GeV.  The $p+p$ results at
the same energy are shown by open squares.  The ratio $R_{AA}(p_x)$ for the
multiplicities in Au+Au collisions to that from $p+p$ collisions for the
triggered jet event is shown in the inset.}
\label{dndpx}
\end{figure}

Figure \ref{dndpx} shows $\rho(p_x)$ for events with a trigger momentum,
$4<P_T\equiv p_T^{\rm trig} <6$ GeV/c. For large negative values of 
$p_x$, $\rho(p_x)$ is suppressed
in Au+Au collisions relative to $pp$ results as expected due to the suppression
of away-side jets \cite{star2}. The situation is reversed for $|p_x|<600$
MeV/c, where the spectrum is dominated by the radiated and rescattered
particles. The momentum constraint, Eq. (\ref{eq:pxconstraint}), is satisfied
by an increase of low momentum particles to compensate for the loss of the
away-side jet. The net number of extra particles, $\Delta N$, is higher in the
Au+Au case as would be expected from the induced-radiation and from the
statistical nature of Eq. (\ref{ngluon}) which converts the radiated energy
into particles with momenta characterized by the temperature. The alteration of
the jets is especially apparent in the inset of Fig. \ref{dndpx} which shows
the ratio of the Au+Au and $pp$ results. At large negative $p_x$ the
suppression is approximately a factor of 3, while the enhancement at low $p_x$
is approximately a factor of two.

Also shown in Fig. \ref{dndpx} (open circles) is the distribution of the
particles in Au+Au collisions that stems from the induced radiation described in
Eq. (\ref{elos}). As expected, these particles are at much lower $p_T$ than
those from $pp$ collisions due to their thermalizing with the medium.

For this range of triggers the number of extra particles, both charged and
neutral, produced at mid-rapidity was approximately $\Delta N=19$, versus 14
for $pp$ collisions. We emphasize that this is a relative increase and 
that determining $\rho_0$ is difficult, as discussed later. 
The numbers increase by $\sim 20$\% if all rapidities are
evaluated.  Of course, experiments would detect a reduced number after
accounting for efficiencies and acceptances. By itself, this number provides
important insight into the driving force behind the disappearance of away-side
jets. If the disappearance of away-side jets was largely an initial-state effect, 
e.g., $2\to 1$ gluon fusion processes in the gluon saturation model
\cite{kharzeev}, few extra particles would be created, especially at
mid-rapidity. One would then expect to see a reduction in $\Delta N$ for Au+Au
vs $pp$ rather than the increase found here.

If we assume a perfect detector, the momentum integration described in
Eqs. (\ref{eq:pxconstraint}) and (\ref{constr2}), resulted in 
$\Delta P_x=-4.2$ GeV/c, after
triggering on leading hadrons in the $p_T^{\rm trig}=4-6$ GeV/c range. This
verifies momentum conservation in the procedure. As with the calculation of
$\Delta N$, most of this balancing momentum was found in the mid-rapidity
cut. If the mechanism for jet quenching were an initial-state effect, one would
expect less of the balancing momentum to be found with a rapidity near that of
the triggering hadron.

It would be straight-forward to expand this analysis to include a binning in
rapidity as well as $p_x$. One could then analyze the degree to which a jet
spreads its energy and momentum in rapidity. Another possibility would be to
bin with $p_T$ or $E_T$ which would allow the determination of the extra energy
related to the trigger. The calculations shown here were performed with
approximately $4\times 10^4$ trigger particles. Two-dimensional binnings would
require significantly higher statistics. We would certainly expect
large-acceptance experiments such as STAR or PHENIX to be able to perform a
comparable analysis as to what was demonstrated here if they were able to
measure on the order of $10^5$ events with trigger particles.

Given sufficient statistics, the principal difficulty of these sort of analyses
lies in defining a class of similar events for the purpose of background
subtraction. For instance, if one uses multiplicity at mid-rapidity to define
the events, the quantity $\Delta N$ would always be zero. The means for
determining centrality must be independent of the particles used in the
analysis of $\rho$ and $\rho_0$. Furthermore, it must be sufficiently accurate
so that the observation of the triggering jet does not bias the event towards
being more central. It might be quite possible to quantitatively estimate and
correct for such a bias. Fortunately, these effects do not determine the
extraction of $\Delta P_x$ since the base distribution, $\rho_0$ does not
contribute to $P_x$.

We also investigated the degree to which our predictions are sensitive to the
parameters used in the model. By reducing the medium-induced energy loss by a
factor of two, the net number of extra radiated particles was only reduced from
5.5 to about half that number. The energy loss is found to be rather insensitive to
the parton elastic cross section. Once the medium becomes opaque, the behavior
should be fairly insensitive to details as thermalization should mask any
details of the microscopic physics, and the linear dependence of the induced
multiplicity with the energy loss parameters should dampen. Consequently, with
a much higher initial gluon density, e.g., $dN_g/dy
\approx 1000$ is suggested in some gluon saturation models \cite{eskola}, it is
not clear to what degree the distribution of produced particles would be
increased.

In conclusion, we have shown that an analysis of low $p_T$ observables can
unveil the fate of lost jets at RHIC. A similar analysis of experimental data
could provide irrefutable evidence for establishing final-state jet energy loss
as the explanation for jet suppression. A high-statistics multi-dimensional
analysis would provide quantitative insight into the dynamics and dissipation
of jets in high-density matter.

\bigskip

\begin{acknowledgments} The authors wish to thank Ivan Vitev for explaining the
finer points of medium-induced radiation. This work was supported by the
National Science Foundation under Grant No. PHY-0070818.
\end{acknowledgments}

\end{document}